\documentclass[twocolumn,prl,showpacs,superscriptaddress]{revtex4}

\usepackage{graphicx}
\usepackage{dcolumn}
\usepackage{bm}

\begin{document}

\title{Dirac Composite Fermion - A Particle-Hole Spinor}
\author{Jian Yang}
\email{jyangmay1@yahoo.com}
\altaffiliation{Permanent address: 5431 Chesapeake Place, Sugar Land, TX 77479, USA}
%\affiliation{}
%\date{}

\begin{abstract}

The particle-hole (PH) symmetry at half-filled Landau level requires the relationship between the flux number $N_{\phi}$ and the particle number $N$ on a sphere to be exactly $N_{\phi} - 2(N-1) = 1$. The wave functions of composite fermions with $\frac{1}{2}$ "orbital spin", which contributes to the shift "1" in the $N_{\phi}$ and $N$ relationship, are proposed, shown to be PH symmetric, and validated with exact finite system results. It is shown the many-body composite electron and composite hole wave functions at half-filling can be formed from the two components of the same spinor wave function of a massless Dirac fermion at zero-magnetic field. It is further shown that away from half-filling, the many-body composite electron wave function at filling factor $\nu$ and its PH conjugated composite hole wave function at $1-\nu$ can be formed from the two components of the very same spinor wave functions of a massless Dirac fermion at non-zero magnetic field. This relationship leads to the proposal of a very simple Dirac composite fermion effective field theory $\mathcal{L} = i\bar{\psi}{\gamma}^{\mu}({\partial}_{\mu}+ia_{\mu})\psi+\frac{1}{2\pi}{\epsilon}^{\mu \nu \lambda}A_{\mu}{\partial}_{\nu}a_{\lambda}$, where the two-component Dirac fermion field is a particle-hole spinor field coupled to the same emergent gauge field, with one field component describing the composite electrons and the other describing the PH conjugated composite holes. As such, the density of the Dirac spinor field is the density sum of the composite electron and hole field components, and therefore is equal to the degeneracy of the Lowest Landau level. On the other hand, the charge density coupled to the external magnetic field is the density difference between the composite electron and hole field components, and is therefore neutral at exactly half-filling. It is shown that the proposed particle-hole spinor effective field theory gives essentially the same electromagnetic responses as Son's Dirac composite fermion theory does.

\end{abstract}
 \maketitle

\section{\label{sec:level1}I. INTRODUCTION}

One of the most remarkable results in the area of the quantum Hall effect (QHE) is predicted by Halperin, Lee, and Read (HLR) in a pioneering work some twenty year ago \cite{HLR}, where they proposed that the two-dimensional electrons in a strong magnetic field at exact half filling of the lowest Landau level behave like a compressible Fermi liquid at zero magnetic field. According to the HLR theory, the composite fermions are formed by attaching two flux quanta to each of the original electrons. In the mean field approximation, the attached flux cancels the external magnetic field exactly at half filling, resulting in a Fermi liquid description of the composite fermions.  Despite of its many successes, the HLR theory is not explicit if not completely lacks particle-hole (PH) symmetry \cite{Kivelson} \cite{Son} \cite{Son1}. On the other hand, the two-body interaction Hamiltonian when projected onto the lowest Landau level is invariant by an antiunitary PH transformation at half-filled Landau level, and the finite size numerical results seem to confirm the PH symmetry of the ground state \cite{Rezayi} \cite{Geraedts}. It is worth noting that the PH symmetry on gapped QHE states relevant to the fractional QHE at $\nu = 5/2$ have also been extensively studied \cite{MR} \cite{Levin1} \cite{Lee} \cite{JYang}.

This long-standing PH symmetry challenge of the HLR theory has been recently met with a rather remarkable resolution proposed by Son \cite{Son} \cite{Son1}. In the new picture of Son, the composite fermion is a massless Dirac fermion characterized by a Berry
 phase of $\pi$, with the PH symmetry built in at the outset. The Dirac composite fermion proposal has generated a great new interest in this rather old field \cite{Wang}\cite{Murthy}\cite{Wang1}\cite{Wang2}\cite{Geraedts}\cite{Barkeshli}\cite{Levin}\cite{Potter}, which lend strong support to the correctness of this very insightful proposal.

In this paper, we will start approaching the subject from a microscopic wave function point of view, and conclude with a similar but distinct effective field theory. We will use Haldane’s spherical geometry \cite{Haldane}. This geometry has been used widely as an efficient tool to perform the numerical finite size studies for the bulk property of the quantum Hall system, mainly because it is free of boundary effects. There is another advantage of using the spherical geometry, as was recognized by Wen and Zee \cite{Wen} in their study to relate the so called "shift" to a topological quantum number "orbital spin" of a quantum Hall state. This "orbital spin" induced shift would not be manifested in geometries such as a torus. 

We will begin with the following composite fermion wave function proposed by Rezayi and Read \cite{Rezayi1}, and by the present author \cite{Yang}, to illustrate the PH symmetry problem of the HLR theory
\begin{equation}
\label{RezayiRead}  P^e_{LLL} \prod\limits_{i<j}^{N_e} (u_iv_j-u_jv_i)^{2} det(Y_{l_im_i}(\theta_j,\phi_j)) 
\end{equation}
where $(u_j, v_j)=(\cos(\theta_j/2)e^{i\phi_j/2},\sin(\theta_j/2)e^{-i\phi_/2})$ are the spinor variables describing the coordinates $(\theta_j,\phi_j)$ of $j^{th}$ electron occupying the spherical harmonics function state $ Y_{l_i,m_i}(\theta_j,\phi_j)$ , and $P^e_{LLL}$ is the electron lowest Landau level projection operator. Loosely speaking, this wave function can be regarded as a wave function version of the HLR theory in that the Jastrow factor $\prod\limits_{i<j}^{N_e} (u_iv_j-u_jv_i)^{2}$ is effectively attaching two flux quanta to each electron, and the Slater determinant $det(Y_{l_i,m_i}(\theta_j,\phi_j))$ builts a Fermi sea at zero magnetic field, occupying spherical harmonics function $Y_{l,m}$ state from small to large values of angular momentum $l$. 
It is clear from Eq.(\ref{RezayiRead}), the flux number $N_{\phi}$ and electron number $N_e$ relationship is given by 
\begin{equation}
\label{FluxNRelationWrong}N_{\phi} - 2(N_e-1) = 0
\end{equation}
However, this breaks particle-hole symmetry condition. The correct flux number $N_{\phi}$ and electron number $N$ relationship that satisfies the PH symmetry condition is
\begin{equation}
\label{FluxNRelationCorrect}N_{\phi} - 2(N_e-1) = 1
\end{equation}
One can easily verify this by noting that for a system of $N_{\phi}$ flux number and $N_e$ electrons, the number of empty electron states (or the number of hole states) is $N_{\phi}+1-N_e$ since the lowest Landau level degeneracy is $N_{\phi}+1$. Equating this number of empty electron states $N_{\phi}+1-N_e$ to the electron number $N_e$ will result in Eq.(\ref{FluxNRelationCorrect}).

\section{ II. PH SYMMETRIC WAVE FUNCTIONS WITH $\frac{1}{2}$ "ORBITAL SPIN" AT HALF-FILLING }

By comparing Eq.(\ref{FluxNRelationCorrect}) with Eq.(\ref{FluxNRelationWrong}), a straightforward solution to meet the correct $N_{\phi}$ and $N_e$ relationship is to modify Eq.(\ref{RezayiRead}) by replacing $det(Y_{l_im_i}(\theta_j,\phi_j))$ with $det(Y_{\frac{1}{2},l_i,m_i}(\theta_j,\phi_j))$ 
\begin{equation}
\label{CompositeElectron} {\Psi}_0^{e} =   P^e_{LLL} \prod\limits_{i<j}^{N_e} (u_iv_j-u_jv_i)^{2} det(Y_{\frac{1}{2},l_i,m_i}(\theta_j,\phi_j)) 
\end{equation}
where $Y_{\frac{1}{2},l,m}(\theta,\phi)$ is the monopole harmonics with a unit Dirac monopole charge at the center of the sphere. On the surface Eq.(\ref{CompositeElectron}) appears to imply that attaching two flux quanta due to $\prod\limits_{i<j}^{N_e} (u_iv_j-u_jv_i)^{2}$ to each electron would not cancel the external magnetic field even in the mean field approximation, and instead the composite fermions still experience a non-zero magnetic field generated by a unit Dirac monopole at the center of the sphere. However this "non-zero magnetic field", as a result of the shift "1" on the difference between $N_{\phi}$ and $2(N_e-1)$ depends on the curvature of a curved space, and is vanishing on a flat space such as a torus. Following Wen and Zee\cite{Wen}, we attribute this "shift" of $1$ to a  topological quantum number representing a half-integer $\frac{1}{2}$ "orbital spin" degrees of freedom of the composite electrons. The total flux seen by the composite electron is the sum of the magnetic flux, which is effectively zero at half filling, and the coupling of the "orbital spin" to the curvature of the sphere (spin connection). This $\frac{1}{2}$ "orbital spin" is also consistent with requirement to obtain the correct value for the coefficient of the $q^2$ correction to the Hall conductivity \cite{Levin} through a relation between the Hall viscosity and the "orbital spin"\cite{Read}\cite{Nguyen}.

We propose the composite hole wave function ${\Psi}_0^{h}$ obtained from the complex conjugate of the composite electron wave function ${\Psi}_0^{e}$ 
\begin{equation}
\label{CompositeHole} {\Psi}_0^{h} =   P^h_{LLL} \prod\limits_{i<j}^{N_h} (u^*_iv^*_j-u^*_jv^*_i)^{2} det(Y_{-\frac{1}{2},l_i,m_i}(\theta_j,\phi_j)) 
\end{equation}
where $N_h = N_e$, and $ P^h_{LLL}$ is the composite hole lowest Landau level projection operator. This wave function describes the same half-filled Landau level system in terms of the composite holes. 

To validate the composite wave function  ${\Psi}_0^{e}$, we will present the numerical results of finite size systems of $7$ electrons and $8$ electrons at the half-filled Landau level satisfying Eq.(\ref{FluxNRelationCorrect}) in the spherical geometry. For a system of $N_e = 7$ electrons, the total flux number is $N_{\phi} = 2N_e-1 = 13$. There are $N_{\phi}+1 = 14$ states in the lowest Landau level with angular momentum $l = \frac{N_{\phi}}{2} = \frac{13}{2}$ and $m = -\frac{13}{2}, -\frac{11}{2}, -\frac{9}{2}, {\ldots}, \frac{13}{2}$. Without a loss of generality, we will choose the Hilbert space to be sectors having the smallest value(s) of the total $z$-component angular momentum $L_z$, which is either $L_z = -\frac{1}{2}$ or $L_z = \frac{1}{2}$ for odd number of electrons such as $N_e = 7$. All the states are doubly degenerate, with each state in one sector has a PH symmetric state in the other sector.  On the other hand, the Slater determinant $det(Y_{\frac{1}{2},l_i,m_i})$ in ${\Psi}_0^{e}$ is formed with $6$ electrons occupying $Y_{\frac{1}{2},l_i,m_i}$ states with $(l_i, m_i)$ taking values of
\begin{equation}
\label{SixStates}
\begin{array}{l}
(\frac{1}{2}, -\frac{1}{2}), (\frac{1}{2},\frac{1}{2}) \\
(\frac{3}{2}, -\frac{3}{2}), (\frac{3}{2},-\frac{1}{2}), (\frac{3}{2},\frac{1}{2}), (\frac{3}{2},\frac{3}{2})\\
\end{array}
\end{equation}
and one electron occupying one of the following states with $l_i, m_i$
\begin{equation}
\label{OtherStates}
(\frac{5}{2}, -\frac{5}{2}), (\frac{5}{2},-\frac{3}{2}), (\frac{5}{2},-\frac{1}{2}), (\frac{5}{2},\frac{1}{2}), (\frac{5}{2},\frac{3}{2}), (\frac{5}{2} \frac{5}{2})
\end{equation}
For the sector of total $z$-component angular momentum $L_z = -\frac{1}{2}$, state $(\frac{5}{2},-\frac{1}{2})$ will be occcupied, and for $L_z = \frac{1}{2}$ sector,  state $(\frac{5}{2},\frac{1}{2})$ will be occcupied.

For a system of $N_e = 8$ electrons, the total flux number is $N_{\phi} = 2N_e-1 = 15$. There are $N_{\phi}+1 = 16$ states in the lowest Landau level with angular momentum $l = \frac{N_{\phi}}{2} = \frac{15}{2}$ and $m = -\frac{15}{2}, -\frac{13}{2}, -\frac{1}{2}, {\ldots}, \frac{15}{2}$. Again we will choose the Hilbert space to be a sector with the smallest value of the total $z$-component angular momentum $L_z$, which is $L_z = 0$ for even number of electrons such as $N_e = 8$. In contrast to $N_e = 7$ case, all the states are non-degenerate, and are PH symmetric with themselves. On the other hand, the Slater determinant $det(Y_{\frac{1}{2},l_i,m_i})$ in ${\Psi}_0^{e}$ is formed having $6$ electrons occupying $Y_{\frac{1}{2},l_i,m_i}$ states with $l_i, m_i$ specified in Eq. (\ref{SixStates}), and $2$ electrons occupying states specified in Eq. (\ref{OtherStates}).

In Fig. 1(a), we plot a lower part of the energy spectrum in an arbitrary units of a $(N_{\phi}, N) = (13, 7)$ finite system in the lowest Landau level versus angular momentum L in $L_z = \frac{1}{2}$ sector. The two numbers $0.9991$ and $0.9998$ below the energy bar at $L = 2.5$ are respectively the overlap (the top number $0.9991$) between ${\Psi}_0^{e}$ and the corresponding exact numerical state, and the overlap (bottom number $0.9998$) between the PH conjugate of ${\Psi}_0^{e}$ and $({\Psi}_0^{h})^*$, using identity $(Y_{S,l,m})^* = (-1)^{S+m}Y_{-S,l,-m}$. 
In Fig. 1(b), we plot a lower part of the energy spectrum for $(N_{\phi}, N) = (15, 8)$ finite system in $L_z = 0$ sector. The numbers $(0.9975)$, $(0.9987)$, and $(0.9989)$ right below the three energy bars at $L = 0, 2, 4$ are the overlaps of the $3$ wave functions described by ${\Psi}_0^{e}$ with their corresponding exact numerical wave functions. The numbers $(0.9958)$, $(0.9996)$, and $(0.9991)$ under the same energy bars are the overlaps between the PH conjugate of the $3$ wave functions described by ${\Psi}_0^{e}$ and themselves (or $({\Psi}_0^{h})^*$, as ${\Psi}_0^{e} = ({\Psi}_0^{h})^*$ for even number of electrons). From these results, we conclude that ${\Psi}_0^{e}$ and ${\Psi}_0^{h}$ in Eq. (\ref{CompositeElectron}) and Eq. (\ref{CompositeHole}) provide accurate description for the exact low energy states, and ${\Psi}_0^{e}$ and ${\Psi}_0^{h}$ are PH symmetric with each other in the sense $\Theta {\Psi}_0^{e} = ({\Psi}_0^{h})^*$, where $\Theta$ is the PH conjugate operator.

\begin{figure}[tbhp]
%\label{fig:Overlap78e}
%\vspace{0.2cm}
%\includegraphics[width=18cm,height=12cm]{Overlap_Figure78e}
\includegraphics[width=\columnwidth]{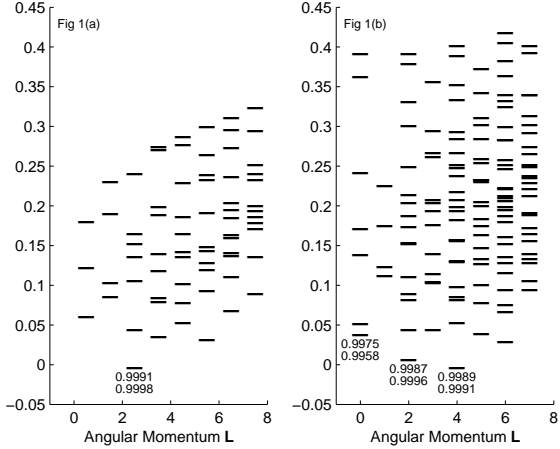}
\caption{\label{fig:Overlap8e} A lower part of the energy spectrum, overlaps (top numbers below energy bars) between ${\Psi}_0^{e}$ and the corresponding exact numerical states, and overlaps (bottom numbers below energy bars) between the PH conjugate of ${\Psi}_0^{e}$ and $({\Psi}_0^{h})^*$, are plotted in Fig. 1(a) for $(N_{\phi}, N_e) = (13,7)$ system, and in Fig. 1(b) for $(N_{\phi}, N_e) = (15,8)$ system.} 
\end{figure}

The monopole harmonics function $Y_{-\frac{1}{2},l,m}$ in Eq. (\ref{CompositeHole}) describes a positive charged particle experiencing a negative unit magnetic monopole at the center of the sphere, the corresponding Hamiltonian $H_{-\frac{1}{2}}$ can be obtained from the following Hamiltonian with $S = -\frac{1}{2}$ \cite{Hasebe} 
\begin{equation}
H_{S} = -\frac{1}{2}{Q}_-^{(S+1)}{Q}_+^{(S)}+\frac{1}{2}S
\end{equation}
where the particle mass is set to $1$ for convenience, and the operators $Q$ are defined by
\begin{equation}
\begin{array}{c}
Q_+^{(S)} = \partial_{\theta} -S {\space} \cot\theta + i \frac{1}{\sin {\theta}} \partial_{\phi}\\
Q_-^{(S)} = \partial_{\theta} +S \cot\theta - i \frac{1}{\sin {\theta}} \partial_{\phi}
\end{array}
\end{equation}
When setting $S = -\frac{1}{2}$, we have
\begin{equation}
H_{-\frac{1}{2}} = -\frac{1}{2}{Q}_-^{(\frac{1}{2})}{Q}_+^{(-\frac{1}{2})}-\frac{1}{4}
\end{equation}
On the other hand, the wave function $Y_{\frac{1}{2},l,m}$ describes a negative charged particle experiencing a positive unit magnetic monopole at the center of the sphere, the corresponding Hamiltonian can be obtained from the complex conjugate of $H_{-\frac{1}{2}}$
\begin{equation}
(H_{-\frac{1}{2}})^* = -\frac{1}{2}{Q}_+^{(-\frac{1}{2})}{Q}_-^{(\frac{1}{2})}-\frac{1}{4}
\end{equation}

It is straightforward to show
\begin{equation}
\label{MatrixEquationZeroField}
\left(
\begin{array}{cc}
H_{-\frac{1}{2}} + \frac{1}{4}& 0 \\
0 & (H_{-\frac{1}{2}})^* + \frac{1}{4}
\end{array}\right)\ = \frac{1}{2}(H^{Dirac}_0)^2
\end{equation}
where $H^{Dirac}_0$ is the Hamiltonian of a massless Dirac particle at zero magnetic field $S = 0$
\begin{equation}
\label{DiracHamiltonian}
H^{Dirac}_S = \left(
\begin{array}{cc}
0 & -iQ_-^{(S+\frac{1}{2})} \\
-i{Q}_+^{(S-\frac{1}{2})} & 0
\end{array}\right)\
\end{equation}
This means the original single composite electron wave function and composite hole wave function are identical, with different coordinates,  to the two components of the same wave function of a massless Dirac particle at zero magnetic field
\begin{equation}
\label{ZeroFieldDirac} \left(
\begin{array}{c}
\phi^{h}_{0,l,m}\\
\phi^{e}_{0,l,m}
\end{array}\right)\
= \left(
\begin{array}{c}
Y_{-\frac{1}{2},l,m}\\
- i Y_{\frac{1}{2},l,m}
\end{array}\right)\
\end{equation}
We will only use the positive energy wave functions, as the negative energy wave functions are the same with a sign change for the lower component \cite{Hasebe}.

We can rewrite ${\Psi}_0^{h}$ and ${\Psi}_0^{e}$ into a two-component compact form using the Dirac wave function at zero magnetic field
\begin{equation}
\label{DiracCompositeFermion} \left(
\begin{array}{c}
{\Psi}_0^{h} \\
{\Psi}_0^{e}
\end{array}\right)\
=   \left(
\begin{array}{c}
P^h_{LLL} \prod\limits_{i<j}^{N_h} (u^*_iv^*_j-u^*_jv^*_i)^{2} det(\phi^{h}_{0,l_i,m_i}(\theta_j,\phi_j)) \\
P^e_{LLL} \prod\limits_{i<j}^{N_e} (u_iv_j-u_jv_i)^{2} det(\phi^{e}_{0,l_i,m_i}(\theta_j,\phi_j))
\end{array}\right)\
\end{equation}
where $N_h = N_e$. The key conclusion is that the many-body composite electron and composite hole wave functions at half-filling can be formed from the two components of the same spinor wave function of a massless Dirac fermion at zero-magnetic field. Of course, in Eq. (\ref{DiracCompositeFermion}) the composite electron and composite hole have different coordinates, and the two component form of wave functions should not be considered as the many-body wave functions of the composite Dirac fermions.

\section{ III. PH CONJUGATED WAVE FUNCTIONS AWAY FROM HALF-FILLING}

In this section, we will move away from the half-filling. First let's modify the matrix equation Eq. (\ref{MatrixEquationZeroField}) to the following
\begin{equation}
\label{MatrixEquationNonZeroField}
\begin{array}{l@{}l}

\left(
\begin{array}{cc}
H_{S-\frac{1}{2}} - \frac{1}{2}(S-\frac{1}{2}) & 0 \\
0 & (H_{-S-\frac{1}{2}})^* + \frac{1}{2}(S+\frac{1}{2})
\end{array}\right)\
\\
= \frac{1}{2}(H^{Dirac}_S)^2 
\end{array}
\end{equation}
where the Hamiltonian $H_{S-\frac{1}{2}}$ describes a positive charged particle experiencing $S$ magnetic monopole charge in addition to a negative unit monopole charge due to the $\frac{1}{2}$ "orbital spin", and $(H_{-S-\frac{1}{2}})^*$, formed from the complex conjugate of $H_{S-\frac{1}{2}}$ with the sign of $S$ monopole charge flipped, describes a negative charged particle experiencing the same magnetic field, and $H^{Dirac}_S$ is the Hamiltonian of a massless Dirac particle at non-zero magnetic field of $S$ given in Eq.(\ref{DiracHamiltonian}).
This means the original single composite electron wave function and composite hole wave function at a magnetic field of $S$, are identical to the two components of the very same wave function of a massless Dirac particle at a magnetic field of $S$, with different coordinates and different energies. 

In the presence of a magnetic field, the massless Dirac particle forms zero-energy and positive energy Landau levels (we ignore the negative energy Landau levels for the same reason as before). The $n^{th}$ positive energy Landau level wave function can be written as
\begin{equation}
\label{DiracParticle}\left(
\begin{array}{c}
\phi^{h}_{S,n,m}\\
\phi^{e}_{S,n,m}
\end{array}\right)\
= \left(
\begin{array}{c}
Y_{S-\frac{1}{2},S-\frac{1}{2}+n,m}\\
-i Y_{S+\frac{1}{2},S+\frac{1}{2}+(n-1),m}
\end{array}\right)\
\end{equation}
with degeneracy $2(S+n)$ where $n$ is a positive integer. The zero-energy wave function is 
\begin{equation}
\label{DiracParticleZeroLandauLevel} \left(
\begin{array}{c}
\phi^{h}_{S,0,m}\\
\phi^{e}_{S,0,m}
\end{array}\right)\
= \left(
\begin{array}{c}
Y_{S-\frac{1}{2},S-\frac{1}{2},m}\\
0
\end{array}\right)\
\end{equation}
with the degeneracy  $2S$. 

We extend the two-component wave functions Eq. (\ref{DiracCompositeFermion}) to non-half filling
\begin{equation}
\label{DiracCompositeFermionNonHalf} \left(
\begin{array}{c}
{\Psi}_{S}^h \\
{\Psi}_{S}^e
\end{array}\right)\
=   \left(
\begin{array}{c}
P^h_{LLL} \prod\limits_{i<j}^{N_h} (u^*_iv^*_j-u^*_jv^*_i)^{2} det(\phi^{h}_{S,n_i,m_i}(\theta_j,\phi_j)) \\
P^e_{LLL} \prod\limits_{i<j}^{N_e} (u_iv_j-u_jv_i)^{2} det(\phi^{e}_{S,n_i,m_i}(\theta_j,\phi_j))
\end{array}\right)\
\end{equation}
Again we emphasize the composite electron and composite hole have different coordinates, and the two component form of wave functions Eq. (\ref{DiracCompositeFermion}) should not be considered as the many-body wave functions of the Dirac composite fermions. We can calculate the flux number $N_{\phi}$ from either ${\Psi}_{S}^h$ or from ${\Psi}_{S}^e$. From ${\Psi}_{S}^h$ we have $N_{\phi} = 2(N_h-1) - 2S$, and from ${\Psi}_{S}^e$ we have $N_{\phi} = 2(N_e-1) +2S$. From these two results, we have
\begin{equation}
\label{NhNeDifference}
N_h-N_e = 2S
\end{equation}
and
\begin{equation}
\label{NhNeSum}
N_h+N_e = N_{\phi}+1
\end{equation} 
Since the minimum value of $N_e$ is zero, Eq. (\ref{NhNeDifference}) requires the minimum value of $N_h$ to be equal to $2S$, which is exactly the degeneracy of the zero-energy Dirac fermion Landau level. In fact, since the lower component of the zero-energy Landau level wave function $\phi^{e}_{S,0,m} = 0$, when only the zero-energy Landau level is completely filled by the Dirac fermions, we have $N_e = 0$ and $N_h = 2S$ which is consistent with Eq. (\ref{NhNeDifference}). This describes an empty electron system. On the other hand, Eq. (\ref{NhNeSum}) reflects the fact that the wave functions in Eq.(\ref{DiracCompositeFermionNonHalf}) describe two PH conjugated states in the Lowest Landau level. 

If in addition to fill the zero-energy Landau level which is a minimum requirement to satisfy Eq. (\ref{NhNeDifference}), we also fill $n$ more non-zero energy Landau levels. In this case, we have $2S = \frac{N_h}{n+1}-n$ for ${\Psi}_{S}^h$ composite hole state, and the total flux  $N_{\phi} = 2(N_h-1)+1-\frac{N_h}{n+1}+n$. This corresponds to filling factor $\nu_h = \frac{n+1}{2n+1}$. On the other hand, we have $2S = \frac{N_e}{n}-(n+1)$ for the composite electron state ${\Psi}_{S}^e$ which only fills $n$ non-zero energy Landau levels since $\phi^{e}_{S,0,m} = 0$. The total flux for ${\Psi}_{S}^e$ is then given by $N_{\phi} = 2(N_e-1)+1+\frac{N_e}{n}-(n+1)$, which yields the filling factor $\nu_e = \frac{n}{2n+1}$. In fact, the wave function ${\Psi}_{S}^e$ at $\frac{n}{2n+1}$ and $({\Psi}_S^{h})^*$ at $\frac{n+1}{2n+1}$ in Eq. (\ref{DiracCompositeFermionNonHalf}) are respectively identical to Jain's wave functions at the same filling factors \cite{Jain}, which are known to be PH conjugated with each other.

\section{ IV. EFFECTIVE FIELD THEORY}

Based on the results of the previous sections, it is natural to postulate the following effective field theory
\begin{equation}
\label{Lagrangian1}
\mathcal{L} = i\bar{\psi}{\gamma}^{\mu}({\partial}_{\mu}+ia_{\mu})\psi+\frac{1}{2\pi}{\epsilon}^{\mu \nu \lambda}A_{\mu}{\partial}_{\nu}a_{\lambda}
\end{equation}
where the Dirac field $\psi$ is a spinor with upper component describing the composite hole field $\psi_h$, and the lower component describing the composite electron field $\psi_e$
\begin{equation}
\psi = \left(
\begin{array}{c}
\psi_h\\
\psi_e
\end{array}\right)\
\end{equation}
and $a_{\mu}$ is an emergent gauge field, and $A_{\mu}$ is the external electromagnetic field. The term $\frac{1}{2\pi}{\epsilon}^{\mu \nu \lambda}A_{\mu}{\partial}_{\nu}a_{\lambda}$ is required in order to satisfy Eq. (\ref{NhNeSum}) and Eq. (\ref{NhNeDifference}). This can been seen from the equation of motion for $a_0$
\begin{equation}
\label{EquationOfMotion_a0}
\psi^{\dagger}\psi = \psi_h^{\dagger} \psi_h + \psi_e^{\dagger} \psi_e = \frac{B}{2\pi}
\end{equation}
where $B$ is the external magnetic field, which is nothing but Eq. (\ref{NhNeSum}). On the other hand, by differentiating  the action with respect to $A_0$, and equating the result to $\psi_h^{\dagger} \psi_h - \psi_e^{\dagger} \psi_e$ since $\psi_h$ and $\psi_e$ have opposite electromagnetic charges, we can obtain 
\begin{equation}
\label{EquationOfMotion_A0}
\psi_h^{\dagger} \psi_h - \psi_e^{\dagger} \psi_e = \frac{b}{2\pi}
\end{equation}
where $b$ is the emergent magnetic field, which is exactly Eq. (\ref{NhNeDifference}). From Eq. (\ref{EquationOfMotion_a0}) and  Eq. (\ref{EquationOfMotion_A0}), we can obtain
\begin{equation}
\label{HoleDensity}
\psi_h^{\dagger}\psi_h = \frac{B}{4\pi}+\frac{b}{4\pi}
\end{equation}
and
\begin{equation}
\label{ElectronDensity}
\psi_e^{\dagger} \psi_e = \frac{B}{4\pi}-\frac{b}{4\pi}
\end{equation}
and use them to relate the electron filling factor to the composite electron (or hole) filling factor in the same way as was done in Section III.

Finally, we would like to make a few comments on the relationship between Eq.(\ref{Lagrangian1}) and the following effective field theory of Son \cite{Son} \cite{Son1}
\begin{equation}
\label{SonLagrangian}
\mathcal{L}_v = i\bar{\psi}_v{\gamma}^{\mu}({\partial}_{\mu}+ia_{\mu})\psi_v+\frac{1}{4\pi}{\epsilon}^{\mu \nu \lambda}A_{\mu}{\partial}_{\nu}a_{\lambda}+\frac{1}{8\pi}{\epsilon}^{\mu \nu \lambda}A_{\mu}{\partial}_{\nu}A_{\lambda}
\end{equation}
In addition to obviously the same Dirac fermion nature, both theories do not have the PH symmetry breaking Chern-Simons term. Furthermore, they give the same electromagnetic responses. In fact, similar to the charge density equations Eq. (\ref{EquationOfMotion_a0}), Eq. (\ref{HoleDensity}), and Eq. (\ref{ElectronDensity}), one can obtain the current density equations from the equation of motion for $a_i$  
\begin{equation}
\label{EquationOfMotion_a}
j_i = j^e_i + j^h_i = \frac{1}{2\pi}{\epsilon}_{ij}E_j
\end{equation}
by differentiating the action with respect to $A_i$
\begin{equation}
\label{EquationOfMotion_A}
 j^h_i- j^e_i = \frac{1}{2\pi}{\epsilon}_{ij}e_j
\end{equation}
and by combining Eq. (\ref{EquationOfMotion_a}) and Eq. (\ref{EquationOfMotion_A})
\begin{equation}
\label{ElectronCurrent}
j^e_i = \frac{1}{4\pi}{\epsilon}_{ij}(E_j-e_j)
\end{equation}
where  $j^e$ and $j^h$ represent  the current densities  of composite electron field and the composite hole field, $e$ and $E$ are the emergent and external electric fields, and ${\epsilon}_{ij}$ is the antisymmetric unit tensor. Since our Dirac spinor current density given in Eq. (\ref{EquationOfMotion_a}) is twice as large as what is in Son's theory, we can write $j_i = 2{\sigma}^D_{ij}e_j$ where ${\sigma}^D_{ij}$ is Son's Dirac composite fermion conductivity tensor. On the other hand, the electrical current coupled directly to the external electric field given by Eq. (\ref{EquationOfMotion_a}) is exactly the same as obtained from Son's theory Eq. (\ref{SonLagrangian}). As a result, the electron conductivity tensor obtained from Eq. (\ref{EquationOfMotion_a}) and Eq. (\ref{ElectronCurrent}) will be the same as that given by Son's theory \cite{Potter}. 

\section{ V.CONCLUSION}

The PH symmetric wave functions of composite fermions with a $\frac{1}{2}$ "orbital spin" are proposed and validated with exact finite system results. The "orbital spin" plays an key role to relate the composite fermions to the Dirac particles, and to lead to the proposal of a composite particle-hole spinor effective field theory which is shown to give essentially the same electromagnetic responses as Son's Dirac composite fermion theory does.

\end{document}